# Foreign Exchange Market Performance: Evidence from Bivariate Time Series Approach

**Mansooreh Kazemilari [1], Maman Abdurachman Djauhari [2] and Zuhaimy Ismail[3]**

[1,3] Department of Mathematical Sciences, Faculty of Science, Universiti Teknologi Malaysia, UTM, Johor, Malaysia
kmansooreh2@live.utm.my

[2] Institute for Mathematical Research, Universiti Putra Malaysia, 43400 UPM Serdang, Malaysia

**Abstract**

There are many studies dealing with the analysis of similarity among currencies in foreign exchange market by using network analysis approach. In those studies, each currency is represented by a univariate time series of exchange rate return. This is the standard practice to analyze the underlying information in the foreign exchange market. In this paper, Escoufier's $RV$ coefficient is applied to measure the similarity among currencies where each of them is represented by bivariate time series. Based on that coefficient, we analyze the topological structure of the currencies. An example of FOREX analysis will be presented and discussed to illustrate the advantages of $RV$ coefficient.

## 1. Introduction

Global foreign exchange (FX) market is the world's largest and most important financial market, extending over all the countries, completely decentralized, with the highest daily trading volume reaching trillions of US dollars [1]. The FX market's dynamics seems to be more complex than any other market's [2]. Thus, it becomes important to study and comprehend the interactions to capture its pricing mechanism. In this regard, the correlation matrix has long been used to quantify the interactions, and the information generated can be very helpful if enough data has been provided beforehand. However, this information extraction process from the correlation matrix is not as straightforward as it seems [3].
In econophysics, Mantegna in 1999 [4] first proposed network analysis method of minimal spanning tree (MST) to analyze stocks' behavior by applying the correlation coefficient as fundamental input which quantifies the degree of similarity between synchronous time evolutions of a pair of returns. Since the work of Mantegna, MST has become an indispensible tool to filter important information contained in the complex structure of a correlation matrix among stocks and currencies [5-7]. Some examples of this method are done in FX markets and stock market based on correlation in different methods by many authors such as McDonald et al. [5], Brida et al. [8], Ortega and Matesanz [9], Mizuno et al. [10], Naylor et al. [11], Kwapień et al. [12], Keskin et al. [13], Jang et al. [14], Wang et al. [6, 15], Kwapień et al. [2], Sinha and Kovur [16], Tumminello et al. [17] and Bonanno et al. [18]. In all of those studies, currencies are represented by a set of univariate time series and all of the models of FX market and stocks market have focused on the behavior of closing price such as Drożdż et al. [19], Oh et al. [20], Vogel and Saravia [21], Oh et al. [22], Sato and Hołyst [23] and Breymann et al. [24]; losing the possibility of embodying information from bid price, and ask price for currencies.

Nonetheless, in finance, many researchers have used the methods of multivariate statistical analysis, such as factor analysis [25], principal component analysis[26] and cluster analysis[27]. Vector correlation first time has described by Hotelling in 1936 [28]. After that, correlation coefficient between two sets of complex vectors was considered by Masuyama in [29, 30] and Rozeboom in 1965 [31]. Then, the measurement of correlation of vector variables was studied by Kshirsagar [32], Escoufier [33], Coxhead [34], Cramer [35], Schaffer & Gillo [36], Cramer and Nicewander [37], Stephens [38], Ramsay et al. [39], Robert et al. [40] and Roy and Cléroux [41]. The majority of multivariate correlations proposed are actually functions of the canonical correlations. Escoufier [33] and Stephens [38] have introduced a vector coefficient by assuming that two sets of vectors are perfectly correlated if there is an orthogonal transformation that can facilitate their collision. Robert and Escoufier [42] have also shown that the Escoufier coefficient is a unifying tool for some multivariate methods.

Escoufier [33] and Robert and Escoufier [42] introduced the use of the $RV$ coefficient for the presentation of different methods of multivariate analysis. They were presented as solutions of optimization problems under various constraints.

In this paper, we use $RV$ to measure the similarity among multivariate time series. More specifically, we focus our attention on the similarity measure among two-dimensional currencies of bid price and ask price, and then analyze currency's behavior. For that purpose, MST will be used to filter information contained in the complex network of correlation matrix among currencies. As an example of application, 45 major world vector currencies from May 2008 until May 2013 will be studied.

The present research is the first one that shows how the concept of Escoufier's vector correlation can appropriately measure the similarity among multivariate time series in currency network. The motivation of this article is to apply the vector correlation coefficient ($RV$) to define the similarity among currencies where each of them is represented by a multivariate time series, and analyze the vector correlations network in terms of topological structure of the currencies.

## 2. Methodology

In order to measure the similarities and differences in the synchronous time evolution of a pair of currencies, we study the correlation between the daily logarithmic changes in bid price and ask price of two currencies viewed in multi-dimensional setting. Consider the change of the successive differences of the natural logarithm of price, which is defined as:

$$V_i = \ln Z_i(t+1) - \ln Z_i(t), \tag{1}$$

where $Z_i$ is the rate $i$ at the time $t$.

Theoretically, let $X$ be an $(n \times p)$ data matrix (of $p$ variables $X_1, \ldots, X_p$) in $\mathcal{R}^p$ and $Y$ be an $(n \times q)$ data matrix (of $q$ variables $Y_1, \ldots, Y_q$) in $\mathcal{R}^q$ corresponding to two sets of variables defined for the same $n$ observations. A commonly used matrix correlation which allows for a different number of variables is the $RV$-coefficient introduced by Escoufier [33]:

$$RV_{X,Y} = \frac{tr(S_{XY} S_{YX})}{\sqrt{tr(S_{XX}^2) tr(S_{YY}^2)}} \tag{2}$$

where $X$ and $Y$ represent multivariate time series of currencies [33], $S_{XX}$ is the sample covariance of $X$ (resp. $S_{YY}$ is the sample covariance of $Y$) and $S_{XY}$ is cross covariance matrix of the variables $X$ and $Y$. In our study, $X$ and $Y$ are data matrix of bivariate time series of bid price and ask price where $p = q = 2$.

The RV coefficient has the following properties:
- For any $X$ and $Y$ data matrix: $0 \leq RV \leq 1$.

- $RV_{X,Y} = 1$ if $Y = aBX + c$, with $B$ an orthogonal matrix, $a$ a scalar, and $c$ a constant vector.
- $RV_{X,Y} = 0$ where all the variables of one set are uncorrelated to all the variables of the following set; $X^t Y = 0$ or $Y^t X = 0$.
- If $p = q = 1$, then $RV_{X,Y} = r^2$, the square of the univariate correlation (Pearson correlation coefficient).
- If $p = 1$ and $q > 1$, then $RV_{X,Y} = R^2/\sqrt{q}$ where $R^2$ is the determination coefficient when we regress $X$ with respect to $Y$ [43]:

Distance $d$ between two data matrices X and Y (two configurations) can also be defined by using the norm $||A|| = \sqrt{tr(A^t A)}$; $A$ is a square matrix. We define

$$d(X,Y) = \left\| \frac{X^t X}{\sqrt{tr(XX^t)^2}} - \frac{Y^t Y}{\sqrt{tr(YY^t)^2}} \right\| = \left\| \frac{S_{XX}}{\sqrt{tr(S_{XX}^2)}} - \frac{S_{YY}}{\sqrt{tr(S_{YY}^2)}} \right\|$$

$$= \sqrt{2}\sqrt{1 - \frac{tr(S_{XY} S_{YX})}{\sqrt{tr(S_{XX}^2) tr(S_{YY}^2)}}} = \sqrt{2}\sqrt{1 - RV(X,Y)} \qquad (3)$$

Now consider data matrix D of size $(n \times n)$ where each element $d(x,y)$ is a Euclidean distance since it satisfies the following three axioms of metric; (i) $d(x,y) \geq 0$ and $d(x,y) = 0 \Leftrightarrow x = y$, (ii) $d(x,y) = d(y,x)$, and (iii) $d(x,y) \leq d(x,l) + d(l,y)$. This matrix represents dissimilarities network currencies defined by its bid price and ask price. To analyze this network, MST is constructed using Kruskal's algorithm as suggested in [44]. This MST is the principal tool to simplify the complex system of currencies in the form of an optimal tree. MST is helpful for understanding the network structure and properties of FX markets. Beneficial insights on the global structure of the financial data, particularly in the equity and exchange rate markets have been demonstrated with MST methodology [45].

## 3. Data

Data of 45 major world currencies from May 5, 2008 until May 4, 2013 were downloaded from http://www.oanda.com/ for daily bid price and ask price. The foreign exchange rate among currencies is mostly stated in relation to the US dollar. In this study also the daily exchange rate of each country's currency for US dollar is used as basic data.

## 4. Method

To analyze the topological properties of currencies in MST, four major centrality measures (degree, betweeness, closeness and eigenvector centrality) as interpretation tools are examined. Centrality is a basic concept in network analysis [46]. The degree of centrality for a node $i$ refers to the number of links attached to the node $i$. It is defined as:

$$C_D(i) = \deg(i)/(n-1) \tag{4}$$

Degree centrality describes the degree of importance of information for each node according to the idea that more important nodes are more active and therefore should have more connections. The Betweenness centrality describes the frequencies of nodes in the shortest paths between indirectly connected nodes [47]. High centrality scores indicate that a node can reach others on relatively short paths, or that a node lies on considerable fractions of shortest paths connecting others. For node $i$, the betweenness centrality can be computed through the following formula:

$$C_B(i) = \frac{1}{(n-1)(n-2)/2} \sum_{j,k \in G} \frac{d_{(j,k)}(i)}{d_{(j,k)}} \tag{5}$$

where $d_{(j,k)}$ denotes the number of shortest paths between node $j$ and node $k$, and $d_{(j,k)}(i)$ represents the number of shortest paths containing point $i$ as an intermediately in the geodesics between node $i$ and node $k$.

The closeness centrality can be regarded as a measure of the time to spread information from a currency to other reachable currencies in the network, and is defined as the mean geodesic distance (i.e., the shortest path) between a node $i$ and all of the nodes reachable from $i$. [46, 48]. Given a node $i$ and a graph $G$, it can be defined as:

$$C_C(i) = \frac{(n-1)}{\sum_{j \in G} d_G(i,j)} \tag{6}$$

where $d_G(i,j)$ signifies the minimum distance between node $i$ and node $j$. It assesses the effectiveness of one node in a graph. The nodes with higher values are closer to the others (on average)[48, 49].

Eigenvector centrality measure is used to determine which node is connected to most connected nodes. The eigenvector centrality of node $i$ is defined as:

$$e_i = \frac{1}{\lambda_{max}} \sum_{j=1}^{n} (A_{ij} x_j) \quad (7)$$

where $x=(x_1, x_2, ..., x_n)^t$ is the eigenvector associated with the largest eigenvalue $\lambda_{max}$ of the adjacency matrix $A$. It is the weighted average of the scores $x_j$ of all vertices linked to node $i$. We can write Equation (7) in matrix form as: $\lambda_{max} x = Ax$.

The larger the value of $e_i$, the more influence of node $i$ to other nodes directly or indirectly. High-scoring node is the one that has connections to other high-scoring nodes.

## 5. Result and discussion

In this section, to simplify the complex network of multi-dimensional currencies, we apply the similarity matrix based on correlation between multivariate time series of currencies to intuitively understand the network structure. MST tool is applied for 45 currencies. MST is used for filtering networks of financial market that resulting in simpler forms of network that can facilitate the analysis of currency exchange markets [50]. The MST chooses the $n-1$ stronger (i.e. shorter) links which span all the currencies by using Kruskal's algorithm.

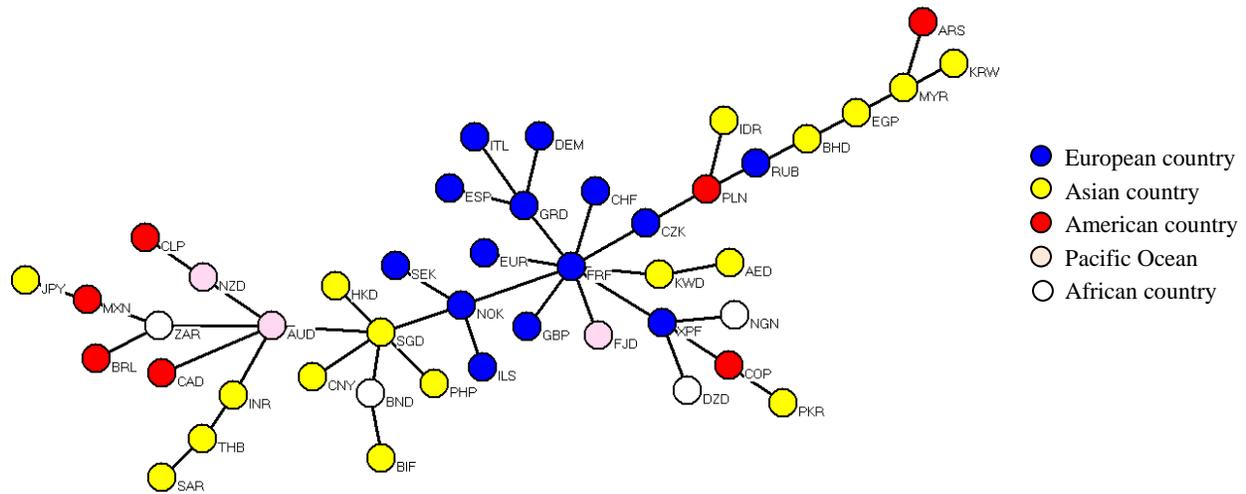

Fig1. MST based on RV coefficient

Fig1 shows several notable results in currency network. We find that most of European corresponding countries are linked together; other currencies are distributed sporadically. FRF (French- Francs) with highest links (9) is in central position of MST while EUR that is a major global currency exist in the lowest linkage. FRF has the highest score in terms of all measures of centralities (degree, betweeness, closeness and eigenvectors). See Table1 for these measures.

Table 1.Countries, respective symbols and centrality measures

| No. | Country | Currency | Degree | Closeness | Betweenness | Eigenvectors |
|---|---|---|---|---|---|---|
| 1 | Algeria | DZD | 0.023 | 0.206 | 0 | 0.081 |
| 2 | Argentina | ARS | 0.023 | 0.116 | 0 | 0.000 |
| 3 | Australia | AUD | **0.114** | 0.251 | 0.396 | 0.110 |
| 4 | Bahamas | BHD | 0.045 | 0.169 | 0.169 | 0.010 |
| 5 | Brazil | BRL | 0.023 | 0.173 | 0 | 0.013 |
| 6 | British | GBP | 0.023 | 0.246 | 0 | 0.189 |
| 7 | Brunei | BND | 0.045 | 0.228 | 0.045 | 0.071 |
| 8 | Burundi | BIF | 0.023 | 0.186 | 0 | 0.022 |
| 9 | Canada | CAD | 0.023 | 0.202 | 0 | 0.034 |
| 10 | France | XPF | 0.091 | 0.257 | 0.174 | **0.266** |
| 11 | Chili | CLP | 0.023 | 0.170 | 0 | 0.011 |
| 12 | China | CNY | 0.023 | 0.226 | 0 | 0.064 |
| 13 | Colombia | COP | 0.045 | 0.208 | 0.045 | 0.090 |
| 14 | Czech | CZK | 0.045 | 0.270 | 0.304 | 0.214 |
| 15 | Egypt | EGP | 0.045 | 0.148 | 0.130 | 0.003 |
| 16 | Europe | EUR | 0.023 | 0.246 | 0 | 0.189 |
| 17 | Fiji | FJD | 0.023 | 0.246 | 0 | 0.189 |
| 18 | France | FRF | **0.205** | **0.324** | **0.743** | **0.617** |
| 19 | Germany | DEM | 0.023 | 0.204 | 0 | 0.080 |
| 20 | Greece | GRD | 0.091 | 0.254 | 0.133 | 0.262 |
| 21 | Hong Kong | HKD | 0.023 | 0.226 | 0 | 0.064 |
| 22 | India | INR | 0.045 | 0.206 | 0.089 | 0.037 |
| 23 | Indonesia | IDR | 0.023 | 0.187 | 0 | 0.025 |
| 24 | Israel | ILS | 0.023 | 0.239 | 0 | 0.095 |
| 25 | Italy | ITL | 0.023 | 0.204 | 0 | 0.080 |
| 26 | Japan | JPY | 0.023 | 0.149 | 0 | 0.004 |
| 27 | Kuwait | KWD | 0.045 | 0.249 | 0.045 | 0.208 |
| 28 | Malaysia | MYR | 0.068 | 0.131 | 0.090 | 0.001 |
| 29 | Mexico | MXN | 0.045 | 0.174 | 0.045 | 0.014 |
| 30 | New Zealand | NZD | 0.045 | 0.204 | 0.045 | 0.037 |
| 31 | Nigeria | NGN | 0.023 | 0.206 | 0.000 | 0.081 |
| 32 | Norway | NOK | 0.091 | **0.312** | **0.539** | **0.311** |
| 33 | Pakistan | PKR | 0.023 | 0.173 | 0 | 0.027 |
| 34 | Philippines | PHP | 0.023 | 0.226 | 0 | 0.064 |
| 35 | Polish | PLN | 0.068 | 0.229 | 0.280 | 0.082 |
| 36 | Russia | RUB | 0.045 | 0.196 | 0.206 | 0.028 |
| 37 | Saudi Arabia | SAR | 0.023 | 0.148 | 0 | 0.004 |
| 38 | Singapore | SGD | **0.136** | 0.289 | 0.541 | 0.209 |
| 39 | South Africa | ZAR | 0.068 | 0.208 | 0.132 | 0.042 |
| 40 | South Korea | KRW | 0.023 | 0.116 | 0 | 0.000 |
| 41 | Spain | ESP | 0.023 | 0.204 | 0 | 0.080 |
| 42 | Sweden | SEK | 0.023 | 0.239 | 0 | 0.095 |
| 43 | Swiss | CHF | 0.023 | 0.246 | 0 | 0.189 |
| 44 | Thailand | THB | 0.045 | 0.173 | 0.045 | 0.013 |
| 45 | Emirate | AED | 0.023 | 0.200 | 0 | 0.064 |

To find the 8 most important currencies based on all four centrality measures, we select 8 nodes of the highest level of each centrality measures and calculate the cumulative distribution for these nodes. Table 2 shows the frequency of these nodes and the level of them in each centrality measures (degree, betweeness, closeness and eigenvector centrality), respectively. We sorted these nodes in terms of the number of frequencies from high to low and presented 8 most important currencies among all centrality measures (See Table 2).

Table2. Frequency and level of 8 most important currencies

| Node | Currency | Frequency | Level |
|---|---|---|---|
| FRF | French -francs | 4 | 1 |
| SGD | Singapore dollar | 4 | 2,2,3,6 |
| NOK | Norwegian kroner | 4 | 6,3,2,2 |
| XPF | CFP France | 3 | 4,8,5,3 |
| CZK | Czech koruna | 3 | 5,10,6,4 |
| GRD | Greece drachma | 3 | 13,5,4,5 |
| AUD | Australian dollar | 2 | 3,4,7,12 |
| PLN | Polish zloty | 1 | 8 |

Table 2 indicates that FRF, SGD and NOK are in the highest levels of four centrality measures with highest frequencies. After that XPF, CZK, GRD, AUD and PLN are in the second level of importance. These currencies are the strongest and most influential currencies. As summary, these currencies have more connections with the others; they are able to control the distribution of information in the network and are able to spread information quickly to the others.

According to the four centrality measures based on the score of each currency, 10 currencies are appeared in the lowest level, namely: ARS, KRW, SAR, JPY, CLP, BRL, PKR, BIF, IDR and CAD. These currencies should be paid more attention by the investors.

## 6. Conclusion

In this paper, we measured the similarity among multi-dimensional currencies of bid price and ask price to find some facts hidden in the currency market of 45 major currencies from 2008-2013 by using $RV$ correlation coefficient. A complex system of bi-dimensional currencies can be described as a network. It efficiently helps us to understand the relationship among currencies. MST is applied to filter important information contained in the complex networks of currencies.

To analyze the statistical properties of currency network, four major centrality measures (degree, betweeness , closeness and eigenvector centrality) as interpretation tools are examined. We presented 8 most influential currencies in terms of the highest scores in all centrality measures.

Previous works used PCC (Pearson Correlation Coefficient) to measure similarity properties of univariate time series of close price. An important contribution of this study is that we combined the method of $RV$ coefficient used for the exploration of bi-dimensional information hidden in currency price and correlation network-based method of MST to investigate the statistical properties of the FOREX.

Our analysis based on the two methods of $RV$ coefficient and MST can be employed to analyze the statistical properties of other financial markets such as stock market, commodity markets, and equity markets. With the growing frequency of financial crises, our work can suggest basic ideas about the relationship between financial crises, currency crises, and the properties of the currency network in terms of multi-dimensional information. The $RV$ coefficient method also can be combined with other network analysis approaches to study the topology of the networks.

## Acknowledgement

The authors gratefully acknowledge the Editor and anonymous referees for their valuable comments and suggestions that led to the final presentation of this work. Special thanks go to (i) the Ministry of Higher Education of Malaysia for financial support, (ii) Universiti Teknologi Malaysia where the first author has initiated this research.